\documentclass[a4paper,11pt]{article}
\usepackage[utf8]{inputenc}
\usepackage[T1]{fontenc}
\usepackage{amsmath}
\usepackage{amssymb}
\usepackage{amsfonts}
\usepackage{graphicx}
\usepackage[margin=1in]{geometry}
\usepackage{float}
\usepackage{tabularx}
\usepackage{array} 
\usepackage{url}
\usepackage{hyperref}

\title{\LARGE \bfseries A Note on Semantic Diffusion}
\author{
\textbf{Alexander P. Ryjov} \\
Department of Computational Mathematics and Cybernetics \\
Lomonosov Moscow State University, Russia \\
\texttt{ryjov@cs.msu.ru}
\and
\textbf{Alina A. Egorova} \\
Department of Mechanics and Mathematics \\
Lomonosov Moscow State University, Russia \\
\texttt{alina.egorova@math.msu.ru}
}
\date{}

\begin{document}

\maketitle

\begin{abstract}
This paper provides an in-depth examination of the concept of \emph{semantic diffusion} as a complementary instrument to large language models (LLMs) for design applications. Conventional LLMs and diffusion models fail to induce a convergent, iterative refinement process: each invocation of the diffusion mechanism spawns a new stochastic cycle, so successive outputs do not relate to prior ones and convergence toward a desired design is not guaranteed. The proposed hybrid framework—“LLM + semantic diffusion”—resolves this limitation by enforcing an approximately convergent search procedure, thereby formally addressing the problem of localized design refinement.

\end{abstract}

\section{Background}

Semantic diffusion was introduced in ~\cite{ref1} as a mechanism that complements large language models (LLMs) in iterative design refinement tasks. It can be viewed as a specific instance of the more general notion of an \emph{adaptive semantic layer}.

The latter was first proposed by Lyapin and Ryjov~\cite{ref2} within the DISNA project~\cite{ref3} to extend search capabilities in large-scale databases. The core idea was to represent user-defined concepts — absent from the database schema — as fuzzy variables whose membership functions enable the retrieval of relevant items.

Initially, these membership functions had to be specified manually by the user, imposing high expertise requirements and thus limiting widespread adoption. Subsequent advances in clustering algorithms (e.g., c-means) enabled the automatic construction of “average” membership functions directly from data; however, such functions reflected only the preferences of a notional “average” user rather than any specific individual.

A pivotal advancement occurred with the introduction of interactive personalization: users could iteratively refine the averaged membership functions by applying fuzzy modifiers (e.g., “slightly cheaper,” “significantly brighter”), thereby tailoring the search process to their unique preferences. This mechanism exemplifies one of the scenarios of hybrid intelligence, combining computational power with intuitive human feedback \cite{ref4}.

Analogous challenges arise in generative design: LLMs and diffusion models, when tasked with producing design drafts, generate each new version independently, preventing gradual, controlled refinement. By integrating LLM-based generation with \emph{semantic diffusion}—an adaptive semantic layer over a local variation space—our approach achieves convergent, user-driven design iteration.

\section{Semantic Diffusion}

The \emph{semantic diffusion} mechanism is engaged once an LLM, using standard prompt‐based generation, produces an initial design that serves as a suitable foundation but requires local refinement. For example, the model might generate a T‐shirt design in which the sleeve needs to be “slightly narrower” and “significantly longer.” The refinement proceeds in three stages:

\begin{enumerate}
  \item \textbf{Construction of a set of local variants.} 
    Let \(P_0\) denote the original value of the selected component parameter \(P\). Define
    \[
      \Delta -\text{(diffusion range, e.g.\ 20\% of }P_0\text{)}, 
      \quad
      \delta - \text{(discretization step, e.g.\ 0.1\% of }P_0\text{)}.
    \]
    Then the set of local variants is
    \[
      \mathcal{V}
      = \bigl\{\,P_0 + k\,\delta \;\bigm|\; k\in\mathbb{Z},\; |k\,\delta|\le\Delta \bigr\}.
    \]
  An example of such diffusion is shown in Fig.~1.

\begin{figure}[ht]
  \centering
  \includegraphics[width=0.9\textwidth]{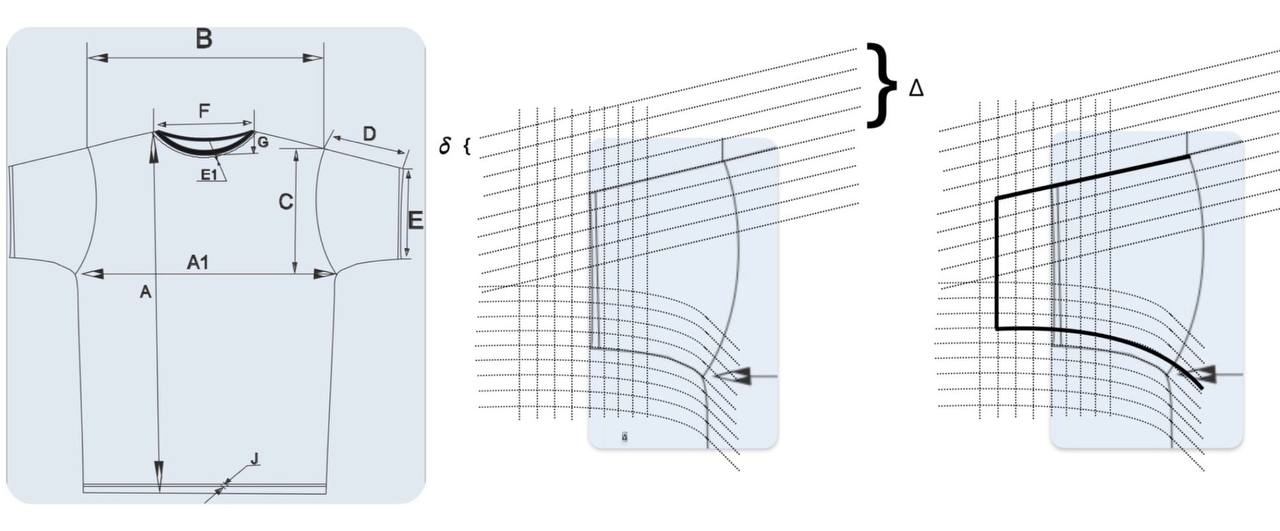}
  \caption{Example of diffusion of the T-shirt sleeve component.}
  \label{fig:diffusion}
\end{figure}

  \item \textbf{Fuzzy modifiers.}
    Navigation through \(\mathcal{V}\) is guided by a \emph{fuzzy modifier}
    \[
      M = (\mathit{Power},\,\mathit{Direction}),
    \]
    where
    \begin{itemize}
      \item \(\mathit{Power}\) denotes the intensity of the change (e.g.\ “slightly,” “moderately,” “significantly”),
      \item \(\mathit{Direction}\)  -- the direction of the modification (e.g., “bigger,” “shorter,” “brighter,” etc., depending on the nature of the parameter being modified).
    \end{itemize}
  
  \item \textbf{Iterative search.}
    The user issues a sequence of modifiers \(M_1, M_2, \dots\), and at each iteration the system selects from \(\mathcal{V}\) the element that best satisfies the current fuzzy criterion. Iterations continue until a satisfactory design is obtained (see Fig.~\ref{fig:diffusion}).

\end{enumerate}

The described mechanism corresponds to the concept of an \emph{adaptive semantic layer} and can be viewed as its particular case: instead of a general-purpose database, the model operates over an instance space with a uniform distribution of objects.

This naturally raises the question of convergence and computational complexity of the associated search procedure. While convergence has been established in the general case of adaptive semantic layers~\cite{ref5}, the more constrained setting of semantic diffusion allows for sharper complexity bounds. Additionally, we propose an extension of the algorithm that accounts for possible user errors—an important consideration in practical applications. These results are discussed in detail in Section~3.

\section{Properties of the Fuzzy Iterative Search Algorithm}

\subsection{Fuzzy Search Algorithms Using Modifiers}

\subsubsection{Problem Formulation}

Let \(N\) objects be uniformly distributed along the interval \([a,b]\), so that adjacent objects are spaced by 
\[
\delta = \frac{|[a,b]|}{N-1}.
\]
Let \(x^*\) be the unknown target point selected by the user, which must be found. Denote the initial position by 
\[
x_0 := 0.
\]
At each iteration the user provides feedback consisting of a direction (“greater”/“less”) and an intensity (“slightly”/“moderately”/“significantly”).

\subsubsection{Membership Function}

Denote the magnitude of the position change at step \(t\) by
\[
\Delta x_t = \lvert x_{t+1} - x_t\rvert,
\]
where \(x_t\) and \(x_{t+1}\) are the positions at times \(t\) and \(t+1\). We construct a membership function \(\mu_p(\Delta x_t)\) to assess the degree of intensity of this change.

\paragraph{``Slightly''}
Define the triangular parameters 
\[
a = 0,\quad b = \Delta_{\min,1},\quad c = \Delta_{\max,1}.
\]
Then
\[
\mu_{\text{slightly}}(\Delta x_t) =
\begin{cases}
0, & \Delta x_t \ge \Delta_{\max,1},\\
\displaystyle \frac{\Delta_{\max,1} - \Delta x_t}{\Delta_{\max,1} - \Delta_{\min,1}}, 
  & \Delta_{\min,1} \le \Delta x_t \le \Delta_{\max,1},\\
1, & \Delta x_t \le \Delta_{\min,1}.
\end{cases}
\]

\paragraph{``Moderately''}
Define
\[
a = \Delta_{\min,2},\quad b = \Delta_{\text{mid},2},\quad c = \Delta_{\max,2}.
\]
Then
\[
\mu_{\text{moderately}}(\Delta x_t) =
\begin{cases}
0, & \Delta x_t \le \Delta_{\min,2}\ \text{or}\ \Delta x_t \ge \Delta_{\max,2},\\
\displaystyle \frac{\Delta x_t - \Delta_{\min,2}}{\Delta_{\text{mid},2} - \Delta_{\min,2}}, 
  & \Delta_{\min,2} \le \Delta x_t \le \Delta_{\text{mid},2},\\
1, & \Delta x_t = \Delta_{\text{mid},2},\\
\displaystyle \frac{\Delta_{\max,2} - \Delta x_t}{\Delta_{\max,2} - \Delta_{\text{mid},2}}, 
  & \Delta_{\text{mid},2} \le \Delta x_t \le \Delta_{\max,2}.
\end{cases}
\]

\paragraph{``Significantly''}
Define
\[
a = \Delta_{\min,3},\quad b = \Delta_{\max,3},\quad c = 1.
\]
Then
\[
\mu_{\text{significantly}}(\Delta x_t) =
\begin{cases}
0, & \Delta x_t \le \Delta_{\min,3},\\
\displaystyle \frac{\Delta x_t - \Delta_{\min,3}}{\Delta_{\max,3} - \Delta_{\min,3}}, 
  & \Delta_{\min,3} \le \Delta x_t \le \Delta_{\max,3},\\
1, & \Delta x_t \ge \Delta_{\max,3}.
\end{cases}
\]

\begin{figure}[H]
  \centering
  \includegraphics[width=0.9\linewidth]{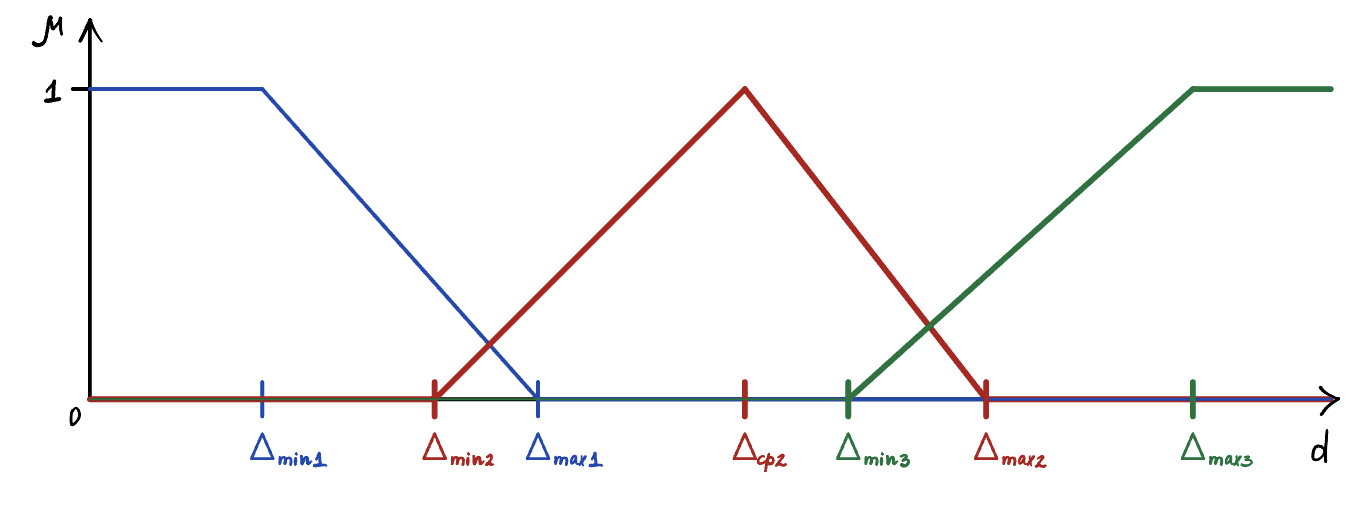}
  \caption{Membership functions for the three intensity levels.}
  \label{fig:membership}
\end{figure}

Each of the three membership functions uses its own thresholds \(\Delta_{\min,p}\), \(\Delta_{\max,p}\) (and \(\Delta_{\text{mid},2}\) for ``moderately'') — see Fig.~\ref{fig:membership}.

\subsection{Algorithm 1: Simple search}

\begin{enumerate}
  \item Set step–size parameters for each intensity level:
    \[
      \text{Slightly:}
      \quad
      \begin{cases}
        \Delta_{\min,1} = \frac{b-a}{k_1},\\
        \Delta_{\max,1} = \frac{b-a}{k_2},
      \end{cases}
      \quad
      \text{with }k_1>k_2\ge2.
    \]
    \[
      \text{Moderately:}
      \quad
      \begin{cases}
        \Delta_{\min,2}    = \frac{b-a}{k_3},\\
        \Delta_{\text{mid},2} = \frac{b-a}{k_4},\\
        \Delta_{\max,2}    = \frac{b-a}{k_5},
      \end{cases}
      \quad
      \text{with }k_3>k_4>k_5\ge2.
    \]
    \[
      \text{Significantly:}
      \quad
      \begin{cases}
        \Delta_{\min,3} = \frac{b-a}{k_6},\\
        \Delta_{\max,3} = \frac{b-a}{k_7},
      \end{cases}
      \quad
      \text{with }k_6>k_7\ge2.
    \]

  \item Initialize the working interval:
    \[
      a = -1,\quad b = 1,\quad x_0 = 0.
    \]
    If \(N\) is the number of objects, then 
    \(\delta = \frac{|[a,b]|}{N-1}\). 
    Choose the precision 
    \(\varepsilon = \frac{\delta}{2} = \frac{|[a,b]|}{2(N-1)}\).

  \item Obtain the first user query \((P,D) = (\text{Power}, \text{Direction})\), denoted \((w_{p,t}, d_t)\), where \(p\in\{1,2,3\}\). Use the corresponding membership function \(\mu_p\) and compute the defuzzified step:
    \[
      \Delta x_t
      = \frac{\displaystyle\int_{\Delta_{\min,p}}^{\Delta_{\max,p}}
               x\,\mu_p(x)\,dx}
             {\displaystyle\int_{\Delta_{\min,p}}^{\Delta_{\max,p}}
               \mu_p(x)\,dx}
      = w_p\,(b-a).
    \]
    In particular:
    \begin{itemize}
      \item \emph{Slightly:} 
        \(\Delta x_t = \frac{b-a}{3}\bigl(\tfrac{7}{2k_1} + \tfrac{1}{k_2}\bigr)
         = w_{\text{slightly}}\,(b-a).\)
      \item \emph{Moderately:} 
        \(\Delta x_t = \frac{b-a}{3}\bigl(\tfrac{1}{k_3} + \tfrac{4}{k_4} + \tfrac{1}{k_5}\bigr)
         = w_{\text{moderately}}\,(b-a).\)
      \item \emph{Significantly:} 
        \(\Delta x_t = \frac{b-a}{3}\bigl(\tfrac{1}{k_6} + \tfrac{2}{k_7}\bigr)
         = w_{\text{significantly}}\,(b-a).\)
    \end{itemize}

  \item Determine the signed change:
    \[
      \operatorname{sgn}(d_t)\,\Delta x_t
      = \begin{cases}
        +\Delta x_t, & d_t = \text{``greater''},\\
        -\Delta x_t, & d_t = \text{``less''}.
      \end{cases}
    \]

  \item Update the position and clamp to \([a,b]\):
    \[
      x_{t+1} = \min\{\,\max\{\,x_t + \operatorname{sgn}(d_t)\,\Delta x_t,\;a\},\;b\}.
    \]

  \item Refine the interval:
    \[
      \begin{aligned}
        a &:= x_t, && \text{if the response is ``greater''},\\
        b &:= x_t, && \text{if the response is ``less''}.
      \end{aligned}
    \]

  \item If \(\lvert b - a\rvert < \varepsilon\), terminate (rounding to the nearest object); otherwise, return to step 3.
\end{enumerate}

The algorithm ends either upon reaching the precision \(\varepsilon\) or when the user confirms the target has been found, yielding \(x_T\) such that
\[
\lvert x_T - x^*\rvert < \varepsilon.
\]

\subsubsection{Convergence}

\textbf{Proposition 1.}  
If the target point \(x^* \in [a_0, b_0]\) and the desired precision \(\epsilon > 0\) are given, then the algorithm guarantees the finding of \(x^*\) with an accuracy no worse than \(\epsilon\) in a finite number of iterations.

\medskip
\textbf{Proof.}  
At each iteration, the boundaries of the current working interval \([a_t, b_t]\) are updated as follows:
\[
a_{t+1} =
\begin{cases}
x_t, & \text{if the direction is ``greater''}, \\
a_t, & \text{if the direction is ``less''};
\end{cases}
\quad
b_{t+1} =
\begin{cases}
b_t, & \text{if the direction is ``greater''}, \\
x_t, & \text{if the direction is ``less''}.
\end{cases}
\]

Thus, the length of the interval at the next step is given by:
\[
|b_{t+1} - a_{t+1}| =
\begin{cases}
b_t - x_t, & \text{if the direction is ``greater''}, \\
x_t - a_t, & \text{if the direction is ``less''}.
\end{cases}
\]

Since by construction \(x_t \in [a_t, b_t]\), it follows that:
\[
|b_{t+1} - a_{t+1}| < |b_t - a_t|.
\]

Moreover,
\[
|b_{t+1} - a_{t+1}| =
\begin{cases}
b_t - x_t = |b_t - a_t|\left(1 - \frac{x_t - a_t}{b_t - a_t}\right), & \text{if ``greater''}, \\
x_t - a_t = |b_t - a_t|\left(1 - \frac{b_t - x_t}{b_t - a_t}\right), & \text{if ``less''}.
\end{cases}
\]

Define the contraction coefficient \(\gamma_t\) as the ratio of the new interval length to the previous one:
\[
\gamma_t =
\begin{cases}
1 - \frac{x_t - a_t}{b_t - a_t}, & \text{if ``greater''}, \\
1 - \frac{b_t - x_t}{b_t - a_t}, & \text{if ``less''}.
\end{cases}
\]
Then,
\[
|b_{t+1} - a_{t+1}| = \gamma_t \cdot |b_t - a_t|,
\]
where \(0 < \gamma_t < 1\), since \(\Delta x_t \in (0, |b_t - a_t|)\).

At each step, the interval length decreases multiplicatively:
\[
|b_t - a_t| = |b_0 - a_0| \cdot \prod_{i=1}^{t} \gamma_i.
\]

As all \(\gamma_i \in (0,1)\), the sequence \(\{|b_t - a_t|\}\) is strictly decreasing and tends to zero as \(t \to \infty\). Therefore, for any \(\epsilon > 0\), there exists \(T\) such that
\[
|b_T - a_T| < \epsilon.
\]
Since \(x^*\) remains within the interval at each step,
\[
x^* \in [a_t, b_t] \quad \forall t,
\]
the target point will be located with the specified accuracy.
\(\square\)

\subsubsection{Algorithm Complexity}

\begin{figure}[h!]
\centering
\includegraphics[width=0.85\linewidth]{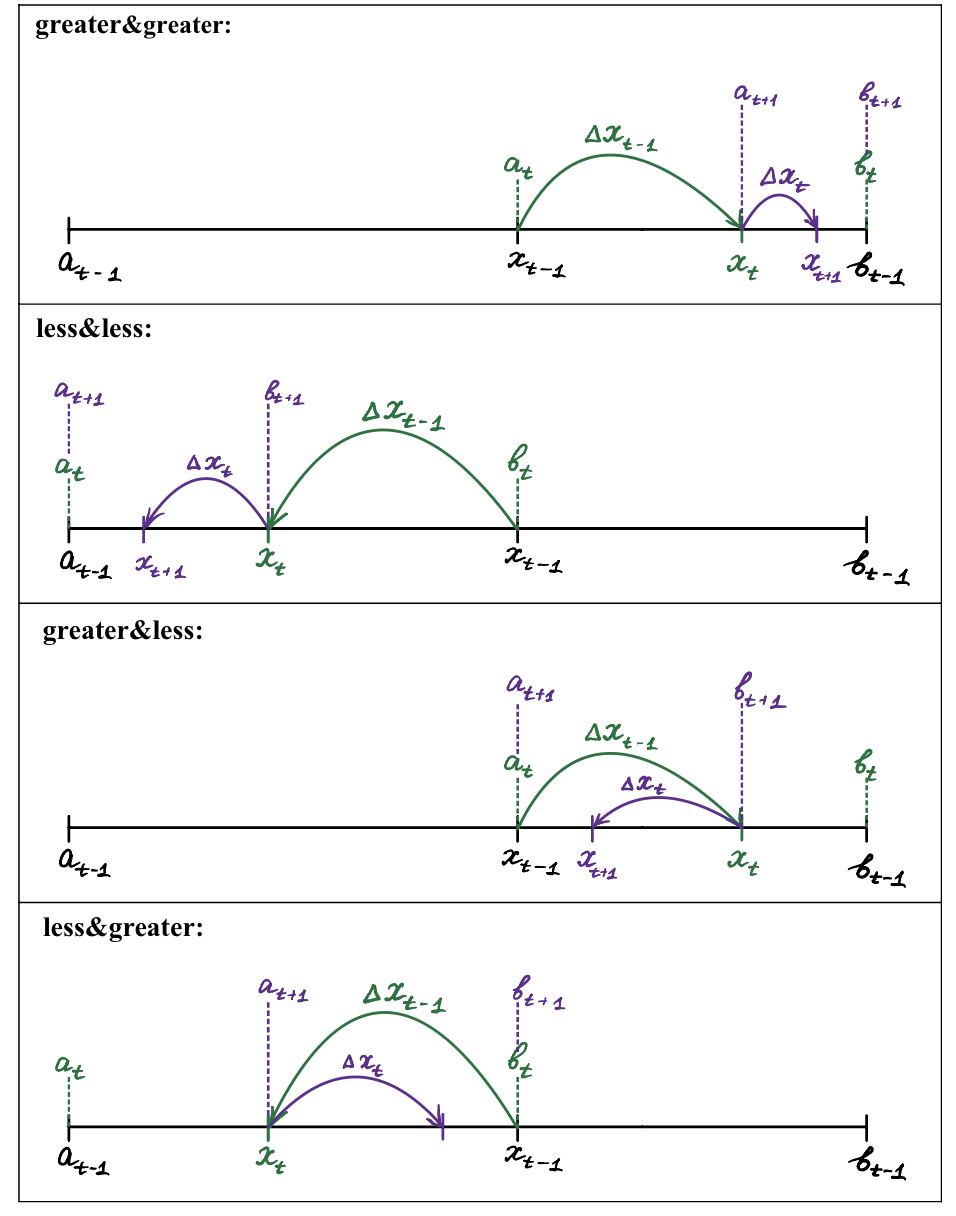}
\caption{Analysis of two consecutive queries.}
\label{fig:cases1}
\end{figure}

Let \( L_t = |b_t - a_t| \) be the length of the current interval \([a_t, b_t]\).  
To estimate the algorithm's complexity, we analyze the number of iterations required to reduce the interval length below the target precision. We consider how the interval changes depending on the last two user responses, taking into account only the direction of modification (see Fig.~\ref{fig:cases1}).

From the figure, we observe:

- For two consecutive identical directions ("greater \& greater" or "less \& less"):
\[
L_{t+1} = L_t - \Delta x_{t-1},
\]

- For alternating directions ("greater \& less" or "less \& greater"):
\[
L_{t+1} = \Delta x_{t-1}.
\]

The step size \(\Delta x_{t-1}\) is given by:
\[
\Delta x_{t-1} = w_{p,t-1} \cdot L_{t-1},
\]
where \(w_{p,t-1}\) is the weight (power) corresponding to the user-specified intensity of change at step \(t-1\).

Substituting \(\Delta x_{t-1}\), we obtain:

- For repeated directions:
\[
L_{t+1} = L_t - w_{p,t-1} \cdot L_{t-1} < L_{t-1} \cdot (1 - w_{p,t-1}).
\]
We can conservatively write:
\[
L_{t+1} = L_{t-1} \cdot (1 - w_{p,t-1}).
\]
This does not affect asymptotic complexity since \(L_t < L_{t-1}\) for all \(t\).

- For alternating directions:
\[
L_{t+1} = w_{p,t-1} \cdot L_{t-1}.
\]

The algorithm terminates when \(L_T < \epsilon\) for some iteration \(T\).

\subsubsection*{Best and Worst Case Analysis}

\begin{enumerate}
  \item \textbf{Best Case.}  
  The interval shrinks as fast as possible: either \(w_{p,t-1} = w_{\text{significantly}}\) with repeated directions, or \(w_{p,t-1} = w_{\text{slightly}}\) with alternating directions.

  In the first best case:
  \[
  L_{t+1} = L_{t-1} \cdot (1 - w_{\text{significantly}})
  \]

  In the second best case:
  \[
  L_{t+1} = w_{\text{slightly}} \cdot L_{t-1}
  \]

  Denote the contraction factor by \(\gamma\), then:
  \[
  L_{t+1} = \gamma \cdot L_{t-1}, \quad \Rightarrow \quad L_t = \gamma^{t/2} \cdot L_0.
  \]
  The termination condition is:
  \[
  \gamma^{t/2} \cdot L_0 < \epsilon,
  \quad \Rightarrow \quad
  t > 2\frac{\log(L_0/\epsilon)}{\log(1/\gamma)}.
  \]

  Hence:
\[
t_{\text{best},1} > 2\frac{\log(L_0/\epsilon)}{\log(1/(1 - w_{\text{significantly}}))}
\]

\[
t_{\text{best},2} > 2\frac{\log(L_0/\epsilon)}{\log(1/w_{\text{slightly}})}
\]

  \item \textbf{Worst Case.}  
  The interval shrinks as slowly as possible: either \(w_{p,t-1} = w_{\text{slightly}}\) with repeated directions, or \(w_{p,t-1} = w_{\text{significantly}}\) with alternating directions.

  In the first worst case:
  \[
  L_{t+1} = L_{t-1} \cdot (1 - w_{\text{slightly}})
  \]

  In the second worst case:
  \[
  L_{t+1} = w_{\text{significantly}} \cdot L_{t-1}
  \]

  Thus:
\[
t_{\text{worst},1} > 2\frac{\log(L_0/\epsilon)}{\log(1/(1 - w_{\text{slightly}}))}
\]

\[
t_{\text{worst},2} > 2\frac{\log(L_0/\epsilon)}{\log(1/w_{\text{significantly}})}
\]
\end{enumerate}

\subsubsection*{Unified Bounds and General Case}

\begin{itemize}
  \item If \(w_{\text{slightly}} = 1 - w_{\text{significantly}}\), then
  \[
  t_{\text{best},1} = t_{\text{best},2} = t_{\text{best}}, \quad
  t_{\text{worst},1} = t_{\text{worst},2} = t_{\text{worst}}.
  \]
  The general case satisfies:
  \begin{equation*}
    t_{\text{worst}} \ge t_{\text{general}} \ge t_{\text{best}},
  \end{equation*}
  \begin{equation*}
    2\frac{\log(L_0/\epsilon)}{\log(1/(1 - w_{\text{slightly}}))}
    \ge t_{\text{general}} \ge
    2\frac{\log(L_0/\epsilon)}{\log(1/w_{\text{slightly}})}.
  \end{equation*}

  \item If \(w_{\text{slightly}} < 1 - w_{\text{significantly}}\), then
  \[
  t_{\text{best},1} > t_{\text{best},2} = t_{\text{best}}, \quad
  t_{\text{worst},2} < t_{\text{worst},1} = t_{\text{worst}}.
  \]
  Hence,
  \begin{equation*}
    t_{\text{worst}} \ge t_{\text{general}} \ge t_{\text{best}},
  \end{equation*}
  \begin{equation*}
    2\frac{\log(L_0/\epsilon)}{\log(1/(1 - w_{\text{slightly}}))}
    \ge t_{\text{general}} \ge
    2\frac{\log(L_0/\epsilon)}{\log(1/w_{\text{slightly}})}.
  \end{equation*}

  \item If \(w_{\text{slightly}} > 1 - w_{\text{significantly}}\), then
  \[
  t_{\text{best},2} > t_{\text{best},1} = t_{\text{best}}, \quad
  t_{\text{worst},1} < t_{\text{worst},2} = t_{\text{worst}}.
  \]
  Hence,
  \begin{equation*}
    t_{\text{worst}} \ge t_{\text{general}} \ge t_{\text{best}},
  \end{equation*}
  \begin{equation*}
    2\frac{\log(L_0/\epsilon)}{\log(1/w_{\text{significantly}})}
    \ge t_{\text{general}} \ge
    2\frac{\log(L_0/\epsilon)}{\log(1/(1 - w_{\text{significantly}}))}.
  \end{equation*}
\end{itemize}

In summary, the general bound can be written as:
\begin{equation*}
  t_{\text{worst}} \ge t_{\text{general}} \ge t_{\text{best}},
\end{equation*}
\begin{equation*}
  2\frac{\log(L_0/\epsilon)}{\log\left(1/\max(w_{\text{significantly}}, 1 - w_{\text{slightly}})\right)}
  \ge t_{\text{general}} \ge
  2\frac{\log(L_0/\epsilon)}{\log\left(1/\min(1 - w_{\text{significantly}}, w_{\text{slightly}})\right)}.
\end{equation*}

Finally, since \(\epsilon = \delta/2 = \frac{|[a,b]|}{2(N-1)}\), the total number of steps can be estimated by:
\begin{equation*}
  2\frac{\log\left(L_0 / \frac{|[a,b]|}{2(N-1)}\right)}{
    \log\left(1/\max(w_{\text{significantly}}, 1 - w_{\text{slightly}})\right)}
  \ge t_{\text{general}} \ge
  2\frac{\log\left(L_0 / \frac{|[a,b]|}{2(N-1)}\right)}{
    \log\left(1/\min(1 - w_{\text{significantly}}, w_{\text{slightly}})\right)}.
\end{equation*}

\subsubsection{Comparison with Binary Search}

This section presents the results of numerical simulations of the fuzzy search algorithm to compare its performance with the classical binary search. For each value of \(N\) (the number of points uniformly distributed on the interval \([-1,1]\)), the optimal values of \(w_{\text{slightly}}, w_{\text{moderately}}, w_{\text{significantly}}\) were selected by brute force to maximize the probability that the fuzzy algorithm finds the target faster than binary search. The win rate is defined as the proportion of target points \(x^*\) for which the number of steps taken by the fuzzy algorithm \(t_{\text{alg}}(x^*)\) is less than that taken by binary search \(t_{\text{bin}}(x^*)\) (see Appendix for implementation details and source code).

The following patterns were observed:
\begin{itemize}
    \item For small values of \(N\) (e.g., \(N=9\)), the optimal coefficients tend to be relatively large, which allows the algorithm to quickly shrink the interval near the boundaries (e.g., \(w_{\text{significantly}} \approx 0.4444\)), resulting in a win rate of approximately 44\%.
    
    \item As the grid becomes denser (\(N=17, 33\)), the optimal values of \(w_{\text{slightly}}\) and \(w_{\text{moderately}}\) decrease (to around \(0.1222\) and \(0.125\), respectively), and \(w_{\text{significantly}}\) also decreases slightly (to around \(0.375\) or \(0.36\)), leading to a win rate of 58--60\%.
    
    \item With further increase in \(N\), a trend emerges: \(w_{\text{slightly}}\) and \(w_{\text{moderately}}\) tend toward zero, while \(w_{\text{significantly}}\) decreases more slowly. This reflects the growing need for finer control when searching in denser grids.
\end{itemize}

The table below shows the optimal \(w\) values, win rates, and distributions of outcomes (wins/draws/losses) for various \(N\) (see Table~\ref{tab:comparison-main}).
\begin{table}[H]
\centering
\small
\begin{tabular}{|c|c|c|c|c|c|}
\hline
\(N\) & \(w_{\text{slightly}}\) & \(w_{\text{moderately}}\) & \(w_{\text{significantly}}\) & Win rate & Wins / Draws / Losses \\
\hline
5   & 0.300 & 0.400 & 0.500 & 20.0\% (1/5)   & 1 / 1 / 3 \\
7   & 0.340 & 0.350 & 0.480 & 28.6\% (2/7)   & 2 / 1 / 4 \\
9   & 0.250 & 0.361 & 0.444 & 44.4\% (4/9)   & 4 / 1 / 4 \\
11  & 0.250 & 0.260 & 0.430 & 36.4\% (4/11)  & 4 / 3 / 4 \\
15  & 0.180 & 0.200 & 0.400 & 46.7\% (7/15)  & 7 / 3 / 5 \\
17  & 0.122 & 0.125 & 0.375 & 58.8\% (10/17) & 10 / 4 / 3 \\
21  & 0.103 & 0.113 & 0.380 & 52.4\% (11/21) & 11 / 5 / 5 \\
25  & 0.086 & 0.096 & 0.372 & 48.0\% (12/25) & 12 / 6 / 7 \\
29  & 0.074 & 0.084 & 0.366 & 51.7\% (15/29) & 15 / 7 / 7 \\
33  & 0.070 & 0.080 & 0.360 & 60.6\% (20/33) & 20 / 5 / 8 \\
37  & 0.065 & 0.075 & 0.355 & 59.5\% (22/37) & 22 / 7 / 8 \\
41  & 0.052 & 0.062 & 0.355 & 58.5\% (24/41) & 24 / 8 / 9 \\
45  & 0.050 & 0.060 & 0.355 & 55.6\% (25/45) & 25 / 6 / 14 \\
51  & 0.045 & 0.055 & 0.350 & 54.9\% (28/51) & 28 / 7 / 16 \\
61  & 0.035 & 0.045 & 0.345 & 52.5\% (32/61) & 32 / 8 / 21 \\
65  & 0.032 & 0.042 & 0.346 & 50.8\% (33/65) & 33 / 11 / 21 \\
73  & 0.029 & 0.039 & 0.344 & 48.0\% (35/73) & 35 / 12 / 26 \\
81  & 0.026 & 0.036 & 0.343 & 46.9\% (38/81) & 38 / 13 / 30 \\
89  & 0.023 & 0.033 & 0.342 & 44.9\% (40/89) & 40 / 14 / 35 \\
97  & 0.022 & 0.032 & 0.341 & 43.3\% (42/97) & 42 / 15 / 40 \\
105 & 0.018 & 0.026 & 0.355 & 47.6\% (50/105) & 50 / 15 / 40 \\
121 & 0.016 & 0.024 & 0.354 & 43.8\% (53/121) & 53 / 17 / 51 \\
129 & 0.015 & 0.023 & 0.354 & 42.6\% (55/129) & 55 / 18 / 56 \\
\hline
\end{tabular}
\vspace{1ex}
\caption{Optimal values of \(w\), win rate, and outcome distributions for various \(N\)}
\label{tab:comparison-main}
\end{table}

\normalsize

As an example, for \(N = 9\), we obtain the following comparison (see Table~\ref{tab:comparison-detailed}):

\begin{table}[H]
\centering
\small
\begin{tabular}{|c|c|c|c|}
\hline
Target position \(x^*\) (index) & \(t_{\text{alg}}(x^*)\) & \(t_{\text{bin}}(x^*)\) & Faster \\
\hline
0 (left endpoint)  & 2 steps  & 3 steps & Fuzzy \\
1                 & 3 steps  & 2 steps & Binary \\
2                 & 3 steps  & 3 steps & Tie \\
3                 & 3 steps  & 4 steps & Fuzzy \\
4 (center, \(x = 0\)) & 4 steps & 4 steps & Tie \\
5                 & 3 steps  & 4 steps & Fuzzy \\
6                 & 3 steps  & 3 steps & Tie \\
7                 & 3 steps  & 2 steps & Binary \\
8 (right endpoint)& 2 steps  & 3 steps & Fuzzy \\
\hline
\end{tabular}
\vspace{1ex}
\caption{Comparison with binary search for \(N = 9\)}
\label{tab:comparison-detailed}
\end{table}

\subsection{Algorithm 2: Search with User Error Tolerance}

We continue to search for the target point $x^*$ selected by the user within the interval $[a, b]$, gradually narrowing it. At each step $t$, the user specifies a direction (``greater'' or ``less'') and a modification intensity $w_{p,t}$ (``slightly,'' ``moderately,'' or ``significantly'').

We modify the algorithm to allow for the possibility of user error, under the assumption that we do not know when the user might make a mistake. However, we assume that the user does not make two consecutive errors, and if they do make a mistake, they try to correct it by using the opposite direction and a stronger modifier. Specifically, we exclude the following cases:

\begin{enumerate}
  \item The user gives two consecutive incorrect directions (e.g., the true point is to the left of the current one, but they choose ``greater'' twice, regardless of the power).
  \item The user first gives an incorrect direction, then a correct one, but in the second input chooses a power that is less than or equal to the previous one (e.g., the point is to the left, and the user inputs ``moderately greater'' followed by ``slightly less''). Exceptions include ``significantly greater'' followed by ``significantly less,'' and vice versa, as the user cannot select a stronger modifier to correct their mistake.
\end{enumerate}

When user error is possible, we must update the interval boundaries differently. If the user gives an incorrect direction, the true point may fall outside the current interval $[a_t, b_t]$ if we proceed as in the previous algorithm. We thus adopt a more cautious approach, reducing the interval more slowly to avoid prematurely excluding possible correct values, even in early stages (e.g., at $x_0 = 0$ with $a_0 = -1$, $b_0 = 1$). We update the interval boundaries only in specific cases of direction and power combinations.

\subsubsection*{Interval Update Rules}

Let us describe the interval update procedure in detail.

Assume we start with the initial interval $[a_0, b_0]$ (e.g., $[-1, 1]$) and initial point $x_0$ (e.g., $x_0 = 0$). At the first step, after receiving the initial direction (``greater'' or ``less''), we compute the next point $x_1$ based on the specified power.

At this point, the working interval remains unchanged: $[a_1, b_1] := [a_0, b_0]$. We update the interval boundaries only after receiving the second user input, in one of the following cases:

\begin{itemize}
  \item After receiving \emph{two consecutive identical} directions, regardless of power.
  \item After receiving \emph{two consecutive different} directions with specific power conditions.
\end{itemize}

\begin{figure}[H]
\centering
\includegraphics[width=0.85\linewidth]{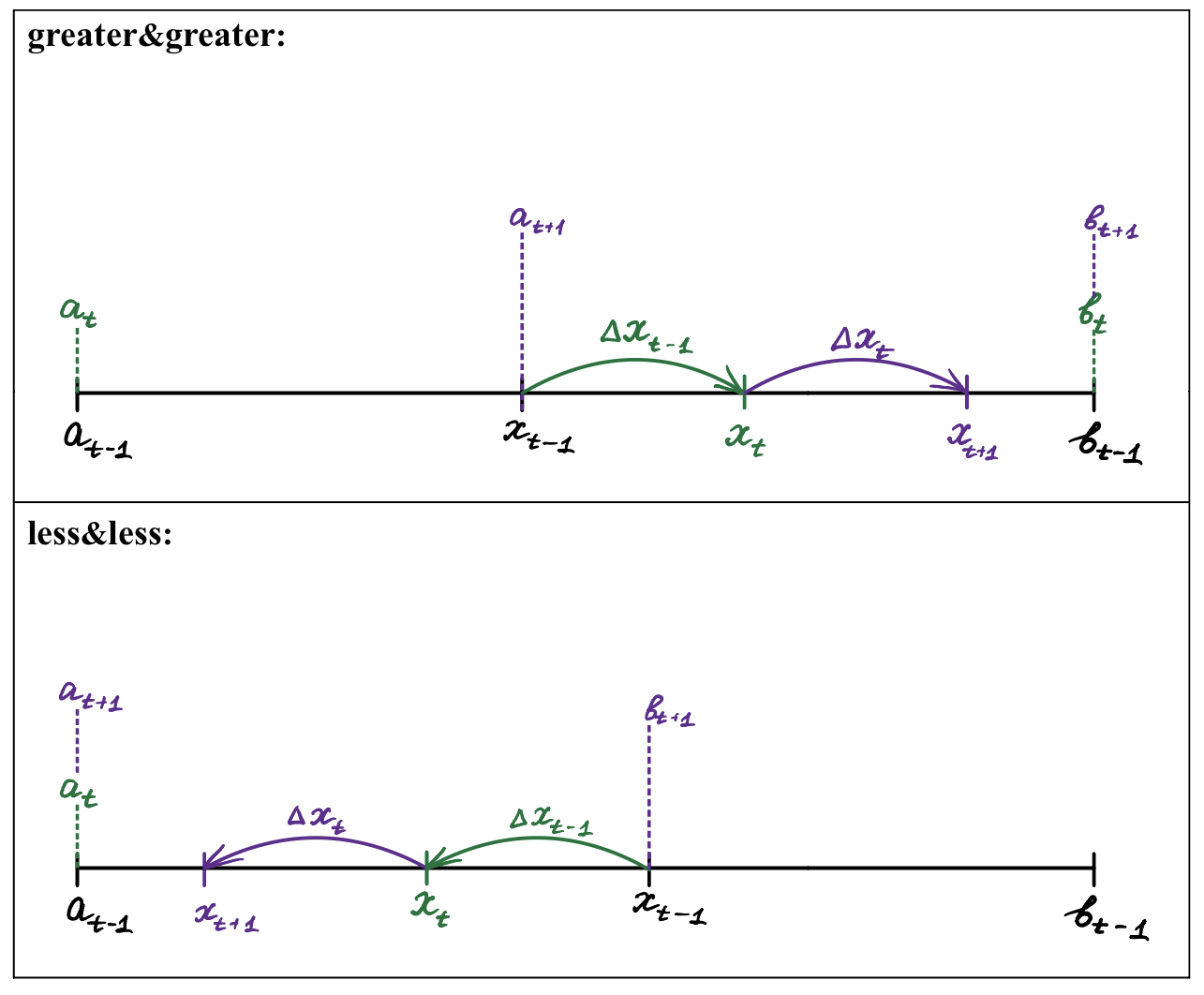}
\caption{Handling two consecutive inputs with the same direction.}
\label{fig:same-dir}
\end{figure}

In the first case (see Fig.~\ref{fig:same-dir}), for example, if the user specifies ``greater'' twice in a row:
\begin{enumerate}
  \item At step 1, we compute $x_1$, but keep the interval unchanged: $[a_1, b_1] := [a_0, b_0]$.
  \item Upon repeating the ``greater'' input, we assume the initial update was likely correct and update the left boundary:
  \[
  a_2 := x_0,
  \]
  while the right boundary remains:
  \[
  b_2 := b_1 = b_0,
  \]
  and we compute $x_2$ accordingly.
\end{enumerate}

Similarly, for two consecutive ``less'' inputs:
\begin{enumerate}
  \item At step 1, compute $x_1$, keep $[a_1, b_1] := [a_0, b_0]$.
  \item Upon repeating ``less,'' update the right boundary:
  \[
  b_2 := x_0,
  \]
  and keep the left boundary:
  \[
  a_2 := a_1 = a_0,
  \]
  then compute $x_2$ using the new interval $[a_2, b_2]$.
\end{enumerate}

In the second case, when the user inputs differ (e.g., first \textit{“greater”}, then \textit{“less”}), if we do not update the interval boundaries, the convergence of the algorithm will be weak—or may not occur at all — when the user continuously alternates directions. Moreover, if we continue to shrink the interval as before, we cannot be confident that the previous steps were error-free, and we risk losing the target point. Therefore, the following approach is proposed:

Let the user input at step $t$ be given by the pair $(w_{p,t}, d_t)$, where $w_{p,t}$ is a fuzzy variable corresponding to the degree of change (\textit{“slightly”}, \textit{“moderately”}, \textit{“significantly”}) and $d_t$ is the direction (\textit{“greater”} or \textit{“less”}).

At step $t-1$, the input is similarly defined: $(w_{p,t-1}, d_{t-1})$.  
\\
The linguistic variables \textit{“slightly”}, \textit{“moderately”}, and \textit{“significantly”} are fuzzy and depend on the user’s perception. Nevertheless, we may assume the existence of a “generalized order” for the degrees of change, which holds at the level of their defuzzified representations. In particular, using the centroid method (see above), we obtain the following numerical estimates:

\[
w_{\text{slightly}} < w_{\text{moderately}} < w_{\text{significantly}}.
\]

Thus, despite possible overlap of the membership functions, the average (centered) values preserve a natural order that corresponds to intuitive notions of intensity. Therefore, if the user specifies the same degree of change for both directions or chooses a weaker one first, it is likely that an error occurred, and we avoid updating the boundaries in such cases.

Let us now examine the first two requests $(w_{p,0}, d_{0})\& (w_{p,1}, d_{1})$ where $d_0 \neq d_1$, i.e., the directions differ:

\begin{itemize}
    \item $d_{0} =$ \textit{greater}; $d_{1} =$ \textit{less}:
    \begin{itemize}
        \item If $w_{p,0} < w_{p,1}$ — the interval remains unchanged, i.e., $[a_2, b_2] := [a_0, b_0]$.
        \item If $w_{p,0} \geq w_{p,1}$ — the interval is reduced based on $x_0$: $[a_2, b_2] := [x_0, b_0]$, \textbf{except for the case “significantly greater” \& “significantly less”}, which is the only case where an error in direction cannot be corrected, since there is no modifier stronger than “significantly”.
    \end{itemize}

    \item $d_{0} =$ \textit{less}; $d_{1} =$ \textit{greater}:
    \begin{itemize}
        \item If $w_{p,0} < w_{p,1}$ — the interval remains unchanged, i.e., $[a_2, b_2] := [a_0, b_0]$.
        \item If $w_{p,0} \geq w_{p,1}$ — the interval is reduced based on $x_0$: $[a_2, b_2] := [a_0, x_0]$, \textbf{except for the case “significantly less” \& “significantly greater”}.
    \end{itemize}
\end{itemize}

This logic is also shown in Table~\ref{tab:direction-change}.

\newcolumntype{C}[1]{>{\centering\arraybackslash}p{#1}}

\begin{table}[H]
\centering
\footnotesize
\begin{tabular}{|C{3.4cm}|C{3.4cm}|C{2.5cm}|C{3cm}|}
\hline
\textbf{First input $(w_{p,0}, d_0)$} & \textbf{Second input $(w_{p,1}, d_1)$} & \textbf{New interval $[a_2, b_2]$} & \textbf{Interval change} \\
\hline
significantly greater & significantly less & $[a_0, b_0]$ & unchanged \\
significantly greater & moderately less & $[x_0, b_0]$ & reduced \\
significantly greater & slightly less & $[x_0, b_0]$ & reduced \\
moderately greater & significantly less & $[a_0, b_0]$ & unchanged \\
moderately greater & moderately less & $[x_0, b_0]$ & reduced \\
moderately greater & slightly less & $[x_0, b_0]$ & reduced \\
slightly greater & significantly less & $[a_0, b_0]$ & unchanged \\
slightly greater & moderately less & $[a_0, b_0]$ & unchanged \\
slightly greater & slightly less & $[x_0, b_0]$ & reduced \\
significantly less & significantly greater & $[a_0, b_0]$ & unchanged \\
significantly less & moderately greater & $[a_0, x_0]$ & reduced \\
significantly less & slightly greater & $[a_0, x_0]$ & reduced \\
moderately less & significantly greater & $[a_0, b_0]$ & unchanged \\
moderately less & moderately greater & $[a_0, x_0]$ & reduced \\
moderately less & slightly greater & $[a_0, x_0]$ & reduced \\
slightly less & significantly greater & $[a_0, b_0]$ & unchanged \\
slightly less & moderately greater & $[a_0, b_0]$ & unchanged \\
slightly less & slightly greater & $[a_0, x_0]$ & reduced \\
\hline
\end{tabular}
\caption{Interval updates depending on the sequence of user inputs}
\label{tab:direction-change}
\end{table}

\begin{figure}[H]
\centering
\includegraphics[width=0.85\linewidth]{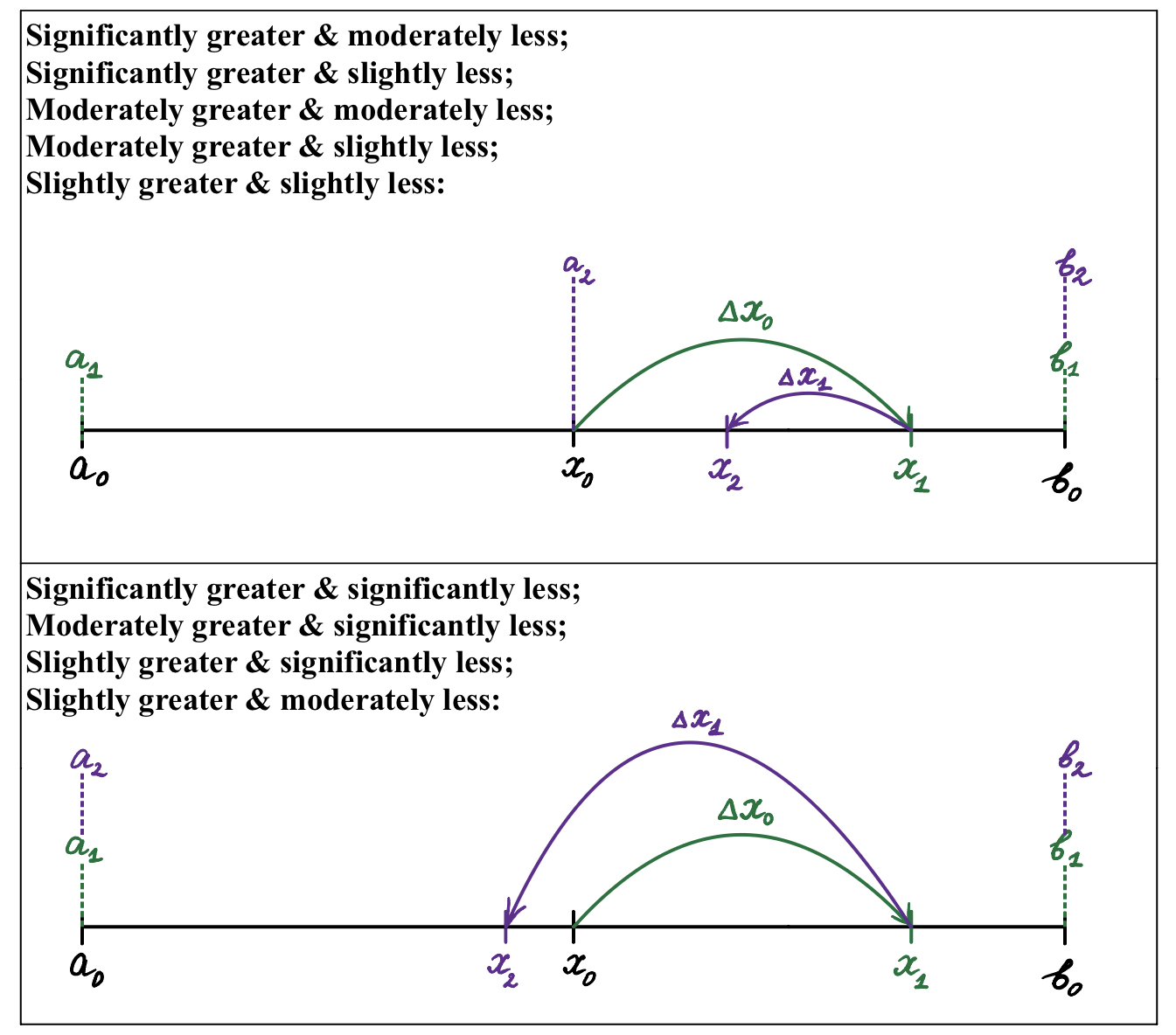}
\caption{Two most recent inputs with opposing directions.}
\label{fig:diff-dir}
\end{figure}

Here, $x_2$ is also computed within the interval $[a_2, b_2]$. For clarity, Figure~\ref{fig:diff-dir} illustrates how the interval boundaries change under two opposing inputs: first \textit{“greater”}, then \textit{“less”}. The reverse case (first \textit{“less”}, then \textit{“greater”}) is constructed analogously. These visualizations only apply to steps 0 and 1; for arbitrary steps $t$ and $t+1$, the scheme becomes invalid, since all previous steps must be taken into account when updating the interval.

\subsubsection{Convergence Analysis of the Search Algorithm under User Errors}

Let the initial interval have boundaries \([a_0, b_0]\) and length
\[
L_0 = b_0 - a_0,
\]
and let the true target point to be found satisfy
\[
x^* \in [a_0,b_0].
\]

The user generates a sequence of queries:
\[
(w_{p,0}, d_{0}),\, (w_{p,1}, d_{1}),\, (w_{p,2}, d_{2}),\, \dots,\, (w_{p,T-1}, d_{T-1}),
\]
where each query contains information about direction and magnitude of adjustment. We assume that the user does not make two consecutive errors and that, in case of an error, they provide a correction in the opposite direction with a stronger modifier (or, if the incorrect input used “significantly,” the next one also uses “significantly” in the opposite direction). Interval updates are performed based on each pair of neighboring queries \( (w_{p,t-1}, d_{t-1}) \& (w_{p,t}, d_{t})\) for \(t\geq 1\). 

Let \(L_t\) denote the length of the working interval after the \(t\)-th update. We assume that interval updates follow a multiplicative principle:
\[
L_{t+1} = L_{t}\cdot M_{t},\quad t\geq 0,
\]
which gives
\[
L_T = L_0 \prod_{t=0}^{T-1} M_t.
\]

The random variable \(M_t\) describes the update effect, depending on the pair of queries \\ \( (w_{p,t-1}, s_{t-1}) \& (w_{p,t}, s_{t})\), and can take one of the following values:
\begin{enumerate}
    \item \textbf{Contracting update:} If the query pair leads to a valid interval narrowing such that the true point \(x^*\) remains inside, then
    \[
    M_t = r_t,\quad 0<r_t<1.
    \]
    Let the probability of this event be \(p_c\).
    
    \item \textbf{Neutral update:} If the query pair results in no interval update:
    \[
    M_t = 1.
    \]
    Let the probability of this event be \(p_n = 1 - p_c\).
\end{enumerate}

\subsubsection*{Probabilistic Description of Updates}

Let \(p_{\text{err}}\) denote the probability that an individual query is erroneous (i.e., an incorrect direction). Each query is correct with probability \(1 - p_{\text{err}}\).

\begin{itemize}
    \item \textbf{Probability of a correct contraction from a query pair:}

    Contraction occurs in the following cases:
\begin{itemize}
    \item Both queries are correct and point in the same direction. The probability is \\
    \(p_\text{same direction}(1 - p_{\text{err}})^2\)
    \item First query is correct, second is incorrect, but both point in the same direction. The probability is \\
    \(p_\text{same direction}(1 - p_{\text{err}})p_{\text{err}}\)
    \item Queries are in opposite directions, but the second power is less than or equal to the first (except for two consecutive “significantly”), and both are correct. The probability is \\
    $p_\text{opposite directions, second power $\leq$ first}(1 - p_{\text{err}})^2$
    \item Same case as above, but the second query is erroneous. The probability is \\
    $p_\text{opposite directions, second power $\leq$  first}(1 - p_{\text{err}})p_{\text{err}}$
\end{itemize}

This yields:

\[
p_c = (p_\text{same direction} + p_\text{opposite directions, second $\leq$ first}) \cdot ((1 - p_{\text{err}})^2 + (1 - p_{\text{err}})p_{\text{err}}) = \pi_c(1 - p_{\text{err}})
\]

where \(\pi_c\) is the conditional probability that a pair of correct queries satisfies the condition for a contracting update (e.g., same direction, or opposing but suitable power relationship).

\item The probability of a neutral update is:
\[
p_n = 1 - p_c.
\]
\end{itemize}

\textbf{Classification note:}  
We have covered all possibilities, so \(p_n = 1 - p_c\). Indeed, the following cases are excluded:
\begin{itemize}
    \item two consecutive erroneous queries;
    \item first query is incorrect, second is correct, and the second power is less than or equal to the first (excluding consecutive “significantly” cases).
\end{itemize}

The remaining cases contribute to neutral updates:
\begin{itemize}
    \item Opposing directions, second power exceeds the first, or two consecutive “significantly”, with both queries correct.
    \item Opposing directions, first correct, second erroneous, second power exceeds first or two “significantly”.
    \item Opposing directions, first erroneous, second correct, second power exceeds first or two “significantly”.
\end{itemize}

Hence, the target point always remains within the interval. We now prove this rigorously via induction.

\textbf{Proposition 2.} If the target point \(x^* \in [a_0, b_0]\), and the user does not make two consecutive errors and always corrects with a stronger opposite-direction query (or repeats “significantly” in the opposite direction), then at every step the point will remain in the current interval.

\textbf{Proof.} By induction.

Base case: by assumption, \(x^* \in [a_0, b_0]\).

Inductive step: assume that for all \(i = 0, \dots, t\), we have
\[
x^* \in [a_i, b_i].
\]
We must show that after step \(t+1\),
\[
x^* \in [a_{t+1}, b_{t+1}].
\]
To do this, we consider not only the next query \((w_{p,t}, d_t)\), but also the two preceding ones that resulted in interval \([a_t, b_t]\) and point \(x_t\):
\[
(w_{p,t-2}, d_{t-2}) \;\&\; (w_{p,t-1}, d_{t-1}).
\]
This is essential to determine the boundary update rule. Without loss of generality, assume \(d_{t-2} =\) “greater” (the “less” case is analogous). We divide all possible situations into three categories:

\begin{enumerate}
    \item At steps \(t-2\) and \(t-1\), the directions match (both “greater”).
    \item At steps \(t-2\) and \(t-1\), the directions are opposite and the pair \((d_{t-2}, d_{t-1})\) results in interval contraction.
    \item At steps \(t-2\) and \(t-1\), the directions are opposite and the pair \((d_{t-2}, d_{t-1})\) does not lead to contraction.
\end{enumerate}

By the inductive assumption:
\[
x^* \in [a_{t-2}, b_{t-2}],\quad x^* \in [a_{t-1}, b_{t-1}].
\]

Now, for each group, we consider cases where \(d_{t-1}\) is incorrect and \(d_t\) is correct, and vice versa. If all three queries are correct, the result is obvious: \(x^* \in [a_{t+1}, b_{t+1}]\) (intervals shrink even more than in Algorithm 1). If the first query is erroneous, we can shift the iteration forward by one, since we will still show that \(x^* \in [a_{t+1}, b_{t+1}]\).

In the diagrams, green arrows indicate correct queries, red arrows indicate errors, and the blue bracket denotes the interval where \(x^*\) lies according to the inductive assumption and the correctness/error pattern. Power indicators: (s) – “significantly,” (m) – “moderately,” (l) – “slightly”; if not specified, any power is acceptable.

\begin{figure}[H]
\centering
\includegraphics[width=0.85\linewidth]{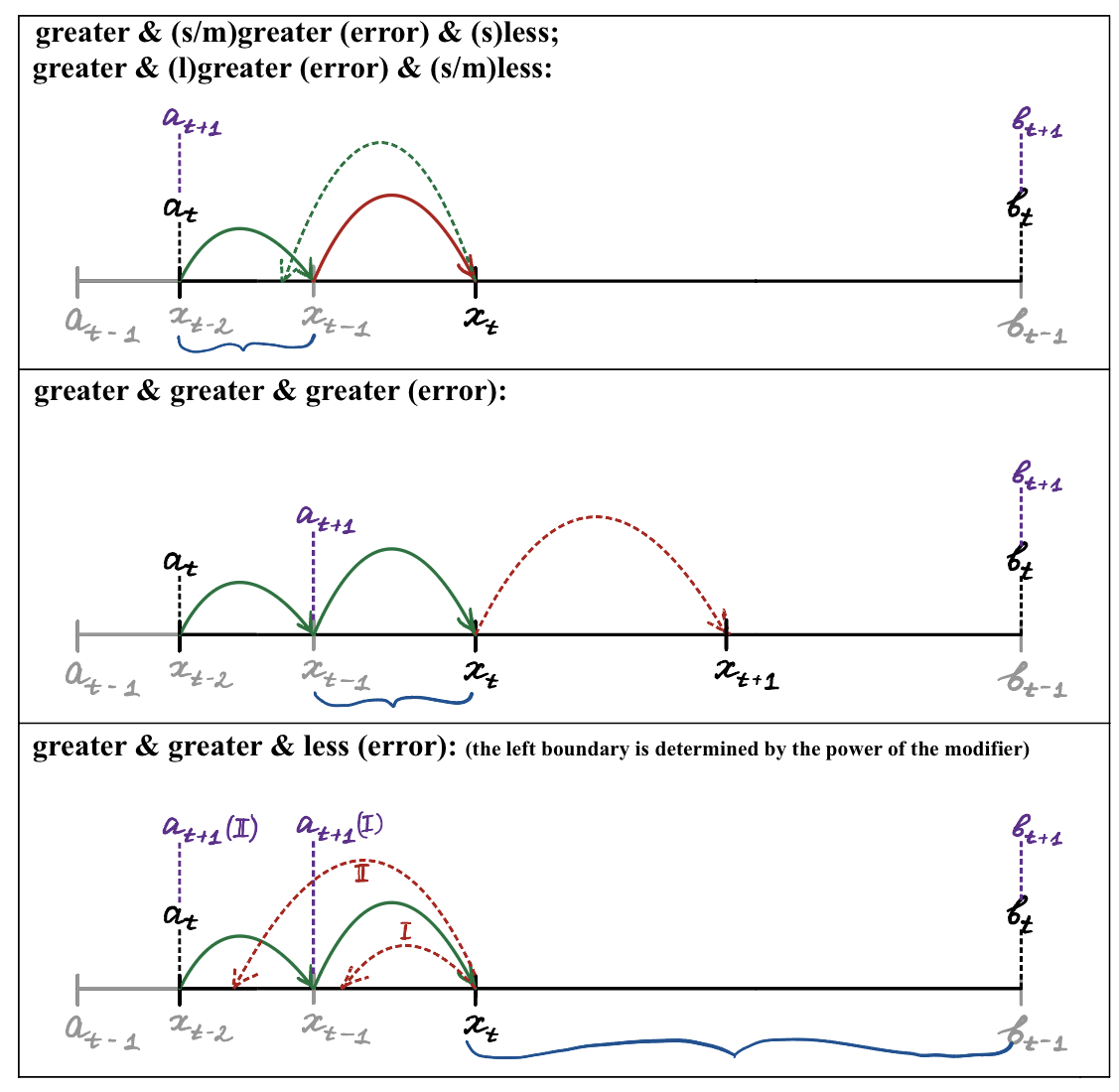}
\caption{Case analysis for Group 1.}
\label{fig:group1}
\end{figure}

\begin{figure}[H]
\centering
\includegraphics[width=0.85\linewidth]{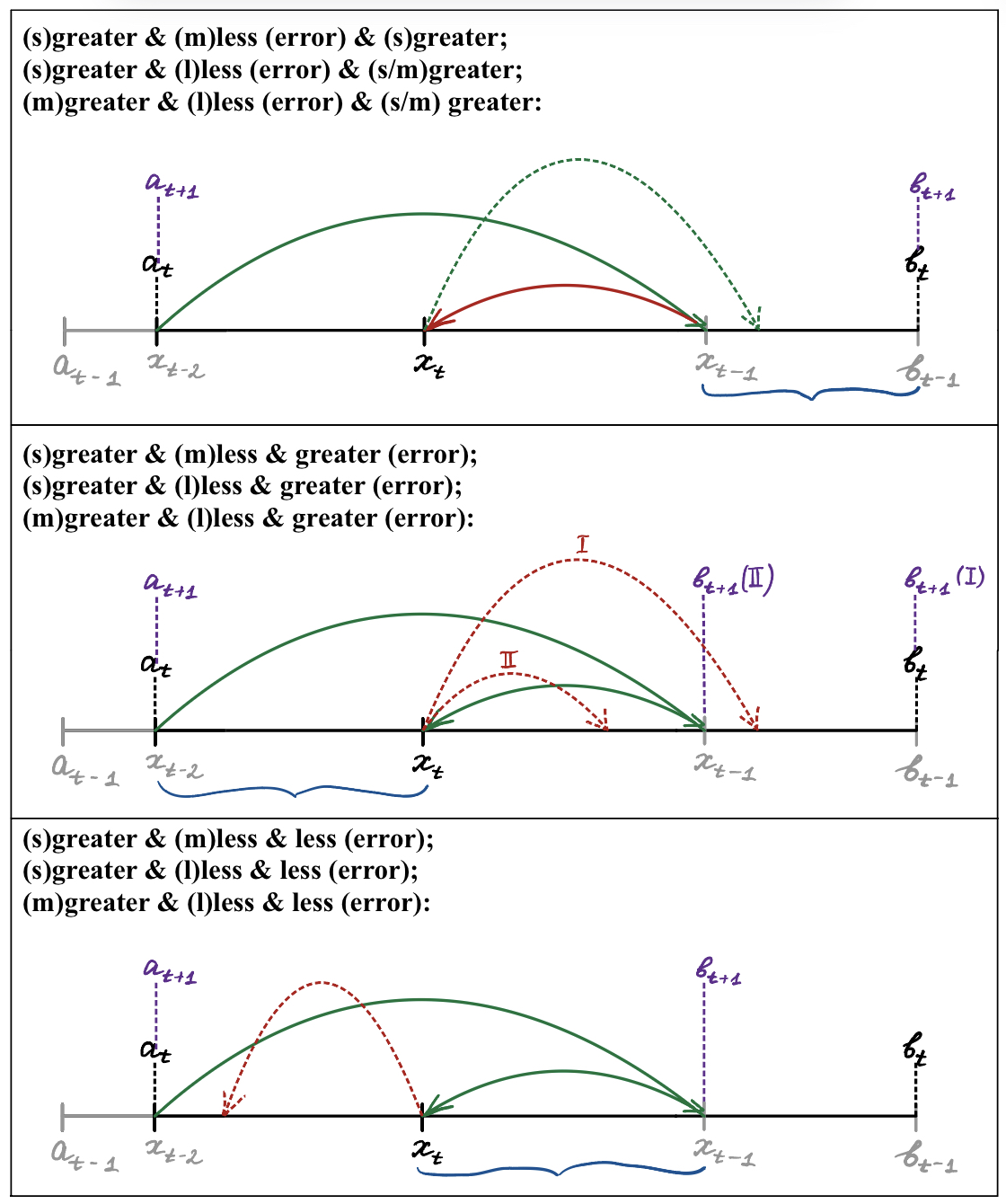}
\caption{Case analysis for Group 2.}
\label{fig:group2}
\end{figure}

\begin{figure}[H]
\centering
\includegraphics[width=0.85\linewidth]{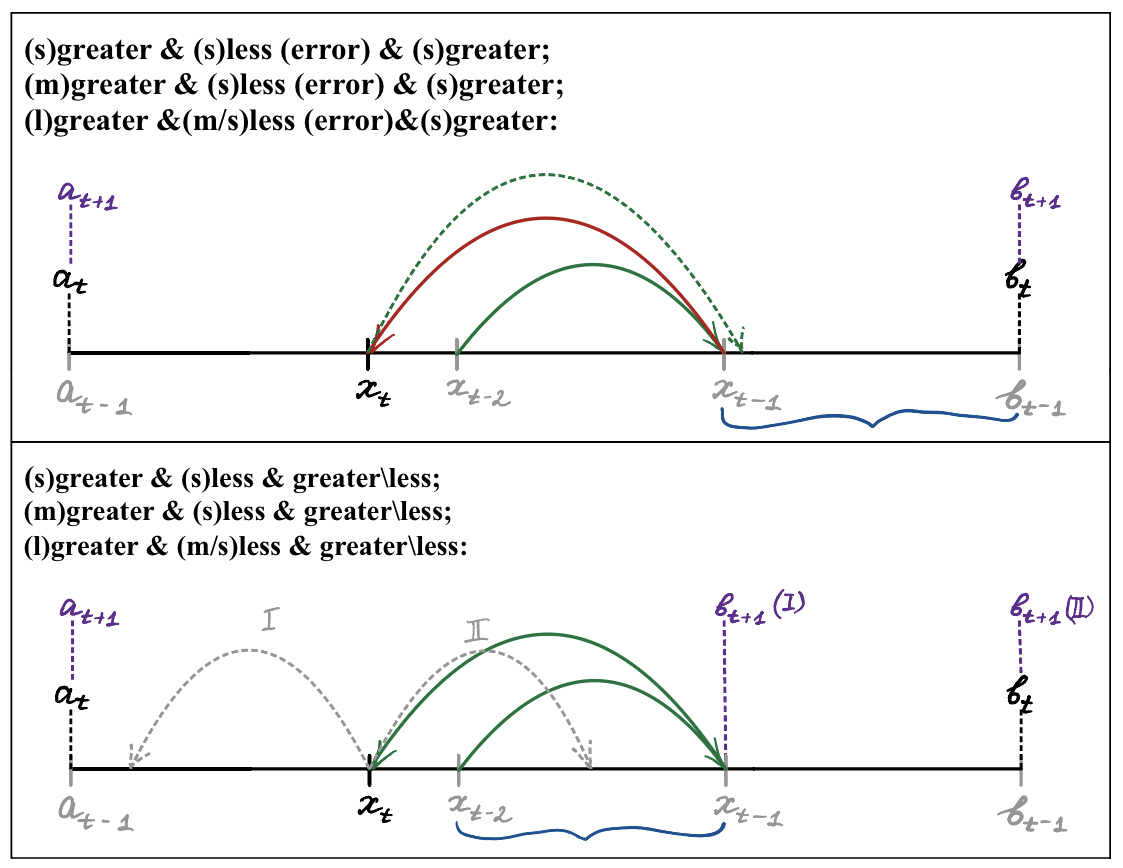}
\caption{Case analysis for Group 3.}
\label{fig:group3}
\end{figure}

In Group 3, only one case with an error remains, since the pair \((w_{p,t-2}, d_{t-2}) \;\&\; (w_{p,t-1}, d_{t-1})\) can either not both be correct, or, if they are indeed correct and the user made a mistake in specifying the power, then regardless of the input, in Case I the direction will coincide with \(d_{t-1}\), or it will be correct in Case II. In either case, the interval containing the target point is preserved.

Thus, we observe that in all possible cases, the interval containing the true point \(x^*\) remains within \([a_{t+1}, b_{t+1}]\). The proposition is proven. \(\quad\square\)

\subsubsection*{Convergence Analysis}

At each step, the interval is either contracted or remains unchanged by construction. We assume that the user does not make more than two consecutive errors, which implies that no more than two consecutive neutral updates can occur (this is the worst-case scenario where the user initially provides a weak modifier, then makes an error by specifying the opposite direction with a stronger modifier—resulting in the first neutral update—and then attempts to return by specifying again the opposite direction with an even stronger (or equal, in the case of “significantly”) modifier). We also assume that the user does not fall into a loop by repeatedly inputting “significantly greater” \& “significantly less”.

The length of the interval after \(T-1\) updates is:
\[
L_T = L_0 \prod_{t=0}^{T-1} M_t.
\]
Taking the natural logarithm gives:
\[
\ln L_T = \ln L_0 + \sum_{t=0}^{T-1} \ln M_t = \ln L_0  + \sum_{t=0}^{T_1} \ln r_t + \sum_{t=0}^{T_2} \ln 1 = \ln L_0  + \sum_{t=0}^{T_1} \ln r_t.
\]

where \(T_1 + T_2 = T - 1\); the first sum corresponds to contractions of the interval with coefficients \(r_t\), and the second corresponds to neutral updates with multiplier 1.

In order for the interval length \(L_T\) to tend to zero, it is necessary and sufficient that:
\[
\sum_{t=1}^{T_1} \ln r_t \;\longrightarrow\; -\infty
\]

Since all \(r_t \leq r_{\max} < 1\), we have \(\ln r_t \leq \ln r_{\max} < 0\).

Furthermore, under the assumption of “no more than two consecutive neutral updates,” in every group of three consecutive steps, there is at least one contracting update. Therefore, in the first \(T\) iterations, the number of contracting updates \(T_1\) satisfies:
\[
T_1 \;\ge\; \Big\lfloor\frac{T - 1}{3}\Big\rfloor.
\]
Hence, using the bound \(r_t \le r_{\max} < 1\), we obtain:
\[
\ln L_T
= \ln L_0 + \sum_{t=1}^{T_1}\ln r_t
\le \ln L_0 + T_1\ln r_{\max}
\le \ln L_0 + \Big\lfloor\frac{T-1}{3}\Big\rfloor\ln r_{\max}
\xrightarrow[T\to\infty]{}\; -\infty,
\]
and therefore,
\[
L_T = e^{\ln L_T} \;\longrightarrow\; 0.
\]

That is, the interval length tends to zero, and the algorithm reaches any desired accuracy \(\varepsilon\).

Thus, we have proven the following:

\textbf{Proposition 3.} If the target point \(x^* \in [a_0, b_0]\) and a desired accuracy \(\epsilon > 0\) is given, then the algorithm guarantees finding \(x^*\) with accuracy no worse than \(\epsilon\) in a finite number of iterations, provided the user does not make two consecutive errors (in the sense of condition 5).\\

\subsubsection{Example of Algorithm Operation with Errors}

Let us consider how the algorithm works in a concrete example. Given:
\begin{itemize}
  \item Interval: \([a,b] = [-1,1]\);
  \item True target point: \(x^* = 0.4\);
  \item Initial position: \(x_0 = 0\);
  \item Number of objects: \(N = 41\), so the spacing between objects is \(\delta = 0.05\);
  \item Modifiers used:
    \begin{itemize}
      \item \(w_\text{significantly} = 0.45\);
      \item \(w_\text{moderately} = 0.35\);
      \item \(w_\text{slightly} = 0.25\).
    \end{itemize}
\end{itemize}

Let us examine the sequence of steps (see Table~\ref{tab:algo2-example}):

\newcolumntype{L}[1]{>{\raggedright\arraybackslash}p{#1}}

\begin{table}[H]
\centering
\footnotesize
\begin{tabularx}{\textwidth}{|c|L{1.25cm}|c|L{2cm}|L{3cm}|L{2.5cm}|X|}
\hline
\textbf{Step} & \textbf{Interval} $[a_t, b_t]$ & $x_t$ & \textbf{User Input} & \textbf{Updated Interval} & $\Delta x$ & $x_{t+1}$ \\
\hline
0 & $[-1, 1]$ & 0 & Moderately greater & $[-1, 1]$ (initial request) & $0.35 \times (1 - (-1)) = 0.7$ & $0 + 0.7 = 0.7$ \\
\hline
1 & $[-1, 1]$ & 0.7 & \textit{Error:} Slightly greater (instead of ``less'') & $[0, 1]$ & $0.25 \times (1 - 0) = 0.25$ & $0.7 + 0.25 = 0.95$ \\
\hline
2 & $[0, 1]$ & 0.95 & Moderately less & $[0, 1]$ (last two inputs did not update) & $0.35 \times (1 - 0) = 0.35$ & $0.95 - 0.35 = 0.6$ \\
\hline
3 & $[0, 1]$ & 0.6 & Slightly less & $[0, 0.95]$ & $0.25 \times (0.95 - 0) = 0.2375$ & $0.6 - 0.2375 = 0.35$ (rounded) \\
\hline
4 & $[0, 0.95]$ & 0.35 & \textit{Error:} Slightly less (instead of ``greater'') & $[0, 0.6]$ & $0.25 \times (0.6 - 0) = 0.15$ & $0.35 - 0.15 = 0.2$ \\
\hline
5 & $[0, 0.6]$ & 0.2 & Moderately greater & $[0, 0.6]$ & $0.35 \times (0.6 - 0) = 0.21$ & $0.2 + 0.21 = 0.41 \approx 0.4$ (rounded) \\
\hline
\end{tabularx}
\vspace{1ex}
\caption{Example 1 of fuzzy search behavior with user errors and interval updates}
\label{tab:algo2-example}
\end{table}

Let us consider another example — the target is \(-0.45\), and the user makes 4 errors (see Table~\ref{tab:fuzzy-errors}):

\newcolumntype{L}[1]{>{\raggedright\arraybackslash}p{#1}}

\begin{table}[H]
\centering
\footnotesize
\begin{tabularx}{\textwidth}{|c|L{1.25cm}|c|L{2cm}|L{3cm}|L{2.5cm}|X|}
\hline
\textbf{Step} & \textbf{Interval} $[a_t, b_t]$ & $x_t$ & \textbf{User Input} & \textbf{Updated Interval} & $\Delta x$ & $x_{t+1}$ \\
\hline
0 & $[-1, 1]$ & 0 & Significantly greater & $[-1, 1]$ (initial request) & $0.45 \times (1 - (-1)) = 0.9$ & $0 - 0.9 = -0.9$ (clipped) \\
\hline
1 & $[-1, 1]$ & $-0.9$ & \textit{Error:} Slightly less (instead of ``greater'') & $[-1, 0]$ & $0.25 \times (0 - (-1)) = 0.25$ & $-0.9 - 0.25 = -1$ (clipped left) \\
\hline
2 & $[-1, 0]$ & $-1$ & Moderately greater & $[-1, 0]$ (unchanged after step 2) & $0.35 \times (0 - (-1)) = 0.35$ & $-1 + 0.35 = -0.65$ \\
\hline
3 & $[-1, 0]$ & $-0.65$ & \textit{Error:} Slightly less (instead of ``greater'') & $[-1, 0]$ (no update due to error) & $0.25 \times (0 - (-1)) = 0.25$ & $-0.65 - 0.25 = -0.9$ \\
\hline
4 & $[-1, 0]$ & $-0.9$ & Moderately greater & $[-1, 0]$ (still unchanged) & $0.35 \times (0 - (-1)) = 0.35$ & $-0.9 + 0.35 = -0.55$ \\
\hline
5 & $[-1, 0]$ & $-0.55$ & \textit{Error:} Slightly less (instead of ``greater'') & $[-1, 0]$ & $0.25 \times (0 - (-1)) = 0.25$ & $-0.55 - 0.225 = -0.775$ \\
\hline
6 & $[-0.9, 0]$ & $-0.775$ & Significantly greater & $[-0.9, 0]$ & $0.45 \times (0 - (-0.9)) = 0.405$ & $-0.775 + 0.405 = -0.37$ \\
\hline
7 & $[-0.9, 0]$ & $-0.37$ & \textit{Error:} Slightly less & $[-0.775, 0]$ & $0.25 \times (0 - (-0.775)) = 0.19375$ & $-0.37 - 0.19375 = -0.175$ \\
\hline
8 & $[-0.775, 0]$ & $-0.175$ & Moderately less & $[-0.775, 0]$ & $0.35 \times (0 - (-0.775)) = 0.27125$ & $-0.175 - 0.27125 = -0.446 \approx -0.45$ \\
\hline
\end{tabularx}
\vspace{1ex}
\caption{Example 2 of fuzzy search behavior with user errors and interval updates}
\label{tab:fuzzy-errors}
\end{table}

\section{Conclusion}

In this paper, we have introduced and formally described the concept of \emph{semantic diffusion}—an adaptive semantic layer that generates a set of local design variants—and developed a fuzzy iterative search algorithm operating in this space. We proved the convergence of the basic algorithm and provided an analysis of its computational complexity. A comparative evaluation against the classical binary search demonstrated that the proposed fuzzy method locates the target elements more quickly or at least as efficiently in a significant number of cases.

Furthermore, we presented an extended version of the algorithm that remains robust in the presence of user errors during interactive refinement. Convergence of this variant was also established, and its enhanced reliability in real-world interactions was confirmed.

Thus, the combination of LLM-based layout generation with subsequent fuzzy search in the semantic diffusion space offers a practical and formally justified approach to the problem of localized design refinement.

\section*{Acknowledgements}

The authors wish to express their deep gratitude to Anwar A. Irmatov for his invaluable comments and recommendations that greatly contributed to the development of this work.

\bibliographystyle{plain}

\end{document}